# Electronic ferroelectricity in monolayer graphene for multifunctional neuromorphic electronics

Le Zhang[1,2], Jing Ding[1,2], Hanxiao Xiang[1,2], Naitian Liu[1,2], Wenqiang Zhou[1,2], Linfeng Wu[1,2], Na Xin[3*], Kenji Watanabe[4], Takashi Taniguchi[5], Shuigang Xu[1,2*]

[1] *Key Laboratory for Quantum Materials of Zhejiang Province, Department of Physics, School of Science, Westlake University, 18 Shilongshan Road, Hangzhou 310024, Zhejiang Province, China*

[2] *Institute of Natural Sciences, Westlake Institute for Advanced Study, 18 Shilongshan Road, Hangzhou 310024, Zhejiang Province, China*

[3] *Department of Chemistry, Zhejiang University, 310058, Hangzhou, China*

[4] *Research Center for Electronic and Optical Materials, National Institute for Materials Science, 1-1 Namiki, Tsukuba 305-0044, Japan*

[5] *Research Center for Materials Nanoarchitectonics, National Institute for Materials Science, 1-1 Namiki, Tsukuba 305-0044, Japan*

[*]Correspondence to: na.xin@zju.edu.cn, xushuigang@westlake.edu.cn

**Abstract:**
Ferroelectricity is intriguing for its spontaneous electric polarization, which is switchable by an external electric field. Expanding ferroelectric materials to two-dimensional limit will provide versatile applications for the development of next-generation nonvolatile devices [1-3]. Conventional ferroelectricity requires the materials consisting of at least two constituent elements associated with polar crystalline structures. Monolayer graphene as an elementary two-dimensional material unlikely exhibits ferroelectric order due to its highly centrosymmetric hexagonal lattices. Nevertheless, two-dimensional moiré superlattices offer a powerful way to engineer diverse electronic orders in non-polar materials [4-7]. Here, we report the observations of electronic ferroelectricity in monolayer graphene by introducing asymmetric moiré superlattice at the graphene/h-BN interface. Utilizing Hall measurements, the electric polarization is identified to stem from electron-hole dipoles, suggesting the electronic dynamics of the observed ferroelectricity. Standard polarization-electric field hysteresis loops, as well as unconventional multiple switchable polarization states, have been achieved. By *in-situ* comparing with control devices, we found that the electronic ferroelectricity in graphene moiré systems is independent of layer number of graphene and the corresponding fine band structures. Furthermore, we demonstrate the applications of this ferroelectric moiré structures in multi-state non-volatile data storage and the emulation of versatile synaptic behaviors, including short-term plasticity, long-term potentiation and long-term depression. This work not only enriches the fundamental understanding of ferroelectricity, but also demonstrates the promising applications of graphene in multi-state memories and neuromorphic computing.

Ferroelectric materials possess electrically switchable spontaneous polarizations, which offer fascinating applications for nonvolatile memories, electric sensors, radio frequency, beyond Boltzmann tyranny transistors, and synaptic devices [1,8-10]. In conventional ferroelectric materials, the spontaneous electric polarization is formed by the spatial separation of the cation and anion, which is switchable by an external electric field ($E$) via small lattice displacements inside a unit cell. Recently, the emergence of two-dimensional ferroelectricity not only provides the opportunity for realizing the miniaturization and multifunction of nonvolatile devices, but also opens the door to the discovery of novel ferroelectricity [11-15]. The reduced dimensionality and designable interlayer stacking in two-dimensional ferroelectric materials enable many unconventional properties different from traditional three-dimensional counterparts [2,3]. Among various two-dimensional ferroelectricity, the interfacial ferroelectricity is particularly intriguing because it arises from stacking non-polar constituents and exhibits high tunability and room-temperature functionality [16-23]. Up to now, there are mainly two kinds of interfacial ferroelectricity discovered in two-dimensional materials: one is the sliding ferroelectricity, where the out-of-plane electric polarization can be switched by in-plane interlayer sliding benefiting from the weak interlayer van der Waals force [17-24]; the other is the unconventional ferroelectricity observed in Bernal-stacked bilayer graphene/h-BN moiré superlattice. In contrast to lattice-driven polarization, the ferroelectricity in bilayer graphene is believed to arise from spontaneous electronic polarization [10,16,25,26], which provides promising applications in ultrafast switchable memories, multiple state storage, and low-power neuromorphic devices [10,27]. Nevertheless, the two-dimensional materials exhibiting electronic ferroelectricity remain extremely rare, limited to bilayer graphene structures with specific alignment with h-BN [16,25,26,28,29].

The origin of unconventional ferroelectricity in bilayer graphene remains elusive. Previous understanding suggested it is highly related to layer-polarized flat moiré bands and tunable quadratic band of bilayer graphene, as evidenced by the accompanying layer-specific anomalous screening effect [16,25,26,30]. The switchable electronic states are presumed to arise from layer polarization of charges and interlayer charge transfer between the top and bottom layers. This interlayer charge transfer model is based on the strong electron-electron interactions in the moiré band.

Here we report the observations of unexpected electronic ferroelectricity in monolayer graphene moiré superlattices, where the layer polarization is essentially absent and linear Dirac band weakens the electron-electron interactions, in contrast to bilayer graphene systems. However, we find that in monolayer graphene, the ferroelectricity as well as gate-specific anomalous screening (GSAS) effect basically resembles that in bilayer graphene. The ferroelectricity in monolayer graphene manifests as the standard polarization-electric field ($P-E$) hysteresis loops and atypical multiple-state switching. Our results argue that layer polarization is not an essential factor for the observation of electronic ferroelectricity in graphene/h-BN superlattices. The underlying mechanism of electronic dynamics in these systems is unveiled by performing a series of $P-E$ loop measurements. Our observations establish graphene as the thinnest ferroelectric material known to exist, enriching the fascinating properties of this wonder material. Furthermore, our discovery will promote the applications of graphene in nonvolatile switchable devices with ultrahigh mobility. To proof this concept, we demonstrate the multifunctionality of multi-state data storages and the emulation of synaptic behaviors, including short-term plasticity (STP), long-term potentiation (LTP), and long-term depression (LTD), by utilizing the emergent ferroelectricity in monolayer graphene.

**Ferroelectric hysteresis**

Our high-quality monolayer graphene devices were made from h-BN encapsulated structures. To create the moiré superlattices, we intentionally aligned the straight edges of graphene with those of both top and bottom h-BN during the assembly process. Raman spectra and second harmonic generation identified the single alignment configuration, namely, graphene crystallographically aligned with the top h-BN and misaligned with the bottom h-BN by 30° (see Methods and Extended Data Fig. 1). The single alignment can be further confirmed from the electronic transport behavior as shown in Fig. 1b, which exhibits typical graphene moiré superlattices hallmark with two satellite peaks at second Dirac point (SDP) besides the main Dirac peak at charge-neutrality point (CNP). The twist angle between graphene and top h-BN is calculated to be 0.68°, resulting in a moiré wavelength of about 12.4 nm (see Methods). The devices were fabricated using dual-gate structures, which allows us to independently tune externally injected total carrier density $n_{\text{total}}$ and the out-of-plane displacement field $D$.

To unveil the layer-dependent ferroelectricity in graphene moiré superlattices, we fabricated a device comprising regions of monolayer, bilayer, and trilayer graphene, which allows us to *in-situ* compare their transport behaviors. The layer number of each region can be easily distinguished from the optical contrast and further confirmed by Raman spectra (see Extended Data Fig. 1). Fig. 1b and 1c show the transport behaviors measured from monolayer and bilayer graphene moiré superlattices, respectively. Compared with bilayer graphene moiré superlattice, the monolayer counterpart exhibits prominent electron-hole asymmetry with hole-side SDP reaching the same order as CNP and weak electron-side SDP, which is consistent with those reported in literatures [31-35]. More characteristics of monolayer graphene can be identified from the nearly $D$-independent CNP (see Extended Data Fig. 2). In the main text, we mainly present the data from monolayer graphene device (Device D1-1), leaving the data from other devices including bilayer, trilayer and twisted bilayer graphene in Extended Data and Supplementary Information. All the data were taken at the temperature $T = 2.2$ K, unless otherwise specified.

Figure 1d and 1e show the transfer characteristics of monolayer graphene device given by sweeping either the top gate ($V_t$) or bottom gate ($V_b$) forward and backward, while keeping the other gate at zero. For $V_b$ sweeps, the forward and backward curves overlap within small gate-sweep ranges ($|V_b|_{\max} \leq 19$ V). However, when $|V_b|_{\max} > 19$ V, four-terminal longitudinal resistance $R_{xx}$ exhibits remarkable hysteresis, which can be easily identified by tracking the positions of CNP and SDP. We can exclude the extrinsic origins of this hysteresis (see Methods). Although both $V_t$ and $V_b$ sweeps show hysteresis, they exhibit different responses across various sweeping ranges. Exceeding a critical $|V_b|_{\max}$, the amplitudes of hysteretic loops measured by the shift of CNP ($\Delta V_b$) increase with the increasing $|V_b|_{\max}$ as shown in the inset of Fig. 1e, while $\Delta V_t$ are almost unchanged in $V_t$ sweep as shown in Fig. 1d. The distinct gate tunability reflects the asymmetric moiré potential at the two interfaces of graphene.

**Gate-specific anomalous screening**

It's worth noting that in Fig. 1d, hole-side SDP does not appear as expected when $|V_t|_{\max} \leq 14$ V

even if the sweep range exceeds the full filling of moiré band (corresponding $\Delta V_t' = 9.7$ V in this device). To further investigate this anomaly, we measured the dual-gate maps of $R_{xx}$ shown in Fig. 2, which strongly depend on the scan directions of both top and bottom gates. For example, when comparing the $V_t$ forward (Fig. 2a) with backward (Fig. 2b) maps, remarkable hysteresis can be observed by tracing the positions of CNP and SDPs. The subtraction between the forward and backward maps is plotted in Fig. 2c. Three sets of hysteretic loops consisting of red and blue peaks are associated with CNP and two SDPs.

Besides the hysteresis behavior, another prominent feature in Fig. 2a is the nearly horizontal lines, manifesting as $R_{xx}$ anomalously independent of $V_t$. In normal dual-gate maps of graphene devices, the resistance peaks associated with CNP and SDPs trace as straight diagonal lines whose slopes are determined by the capacitances of two gate dielectrics, suggesting that both gates can effectively inject carriers and tune the Fermi energy. If the dual-gate maps are converted to $n_{\text{total}} - D$ maps, the diagonal lines turn into vertical lines (see Methods). However, in our device, within some regions (for instance, $-20$ V $< V_t < -5$ V for hole-side SDP in Fig. 2a), the position of SDP abruptly freezes, as if $V_t$ has been screened and does not work anymore. We denote this phenomenon as GSAS. In this device, $V_t$ is the specific gate, stemming from moiré superlattice at the top interface (see Methods). The appearances of GSAS regions depend on the scan direction of fast-scan gate as shown in Fig. 2a and 2b. As the fast-scan gate is changed from $V_t$ to $V_b$ as shown in Fig. 2e and 2f, the GSAS regions tend to shrink, but still exist. Moreover, the slopes of the diagonal line are different before and after GSAS occurs. For instance, if tracing the electron-side SDP in Fig. 2f, the slope of diagonal line is normal when the slow-scan gate ($V_t$) is swept from 25 V to 12 V. Across the region of 12 V $< V_t < 7$ V, electron-side SDP appears as a vertical line indicating $V_t$ is screened. When $V_t < 7$ V, trajectory of electron-side SDP becomes a diagonal line but with a smaller slope compared with initial states ($V_t > 12$ V), indicating that $V_b$ is partially screened in this region. The CNP and hole-side SDP show similar evolution trend, although occur at slightly different regions. The GSAS can be identified more clearly as we plot the corresponding $n_{\text{total}} - D$ maps in Fig. 2d and 2h, where straight lines represent the normal gating effect and oblique lines indicate screening effect. More detailed results show that GSAS also depends on the scan range of fast-scan gate (see Fig. S1 and S2 in Supplementary Information), but independent of the scan direction of slow-scan gate (see Fig. S3 in Supplementary Information).

**Polarization-electric field hysteresis loops**

The presence of GSAS indicates the anomalous external field tunability of charge carrier density in this system. To unveil this, we resort to the Hall measurements. The Hall resistance $R_{xy}$ directly probes the itinerant charge density given by $n_H = -B/eR_{xy}$, where $B$ is external magnetic field, and $e$ is the electron charge. Fig. 3a shows the difference of Hall density ($\Delta n_H$) between forward and backward sweeps as a function of $n_{\text{total}}$ and $E$, in which the fast-scan axis is $E$, achieved by simultaneously sweeping $V_t$ and $V_b$ with a fixed relation (see Methods). In normal dual-gate devices, scanning $E$ will not change $n_H (= n_{\text{total}})$ when fixing $n_{\text{total}}$. Specifically, at $n_{\text{total}} = 0$, we should achieve $n_H = 0$. However, in our device, as shown in Fig. 3b-3f, though $n_{\text{total}}$ is fixed at zero, prominent nonzero polarization $P_{2D} = en_H d_{\text{dipole}} \propto n_H$ (see Methods) can be achieved and is strongly dependent on the scan history of $E$. Similar behavior of tunable $\Delta n_H$ by scanning $E$ at $n_{\text{total}} \neq 0$ can be observed as shown in Fig. 3a. We believe this feature shares common origin as GSAS (see Methods). Normally, when fixing $n_{total} = 0$, $V_t$ and $V_b$ inject holes and electrons (or vice versa) with equal density. In our system, moiré potential at the top interface traps holes (or electrons) injected by $V_t$, which therefore do not contribute to the conductance. As a result, $V_t$ becomes the specific gate for GSAS. One the other hand, because holes (electrons) are localized, itinerant electrons (holes) injected by the other gate ($V_b$) cannot be neutralized leading to the nonzero $n_H$.

Figure 3b shows a typical $P_{2D} - E$ hysteresis loop. Obviously, two distinct states of nonzero remanent polarization ($P_r$) with opposite signs can be achieved at $E = 0$ after a history of applying $E$. The signs of the two states can be reversed by applying a large $E$ with the opposite direction. More generally, the polarization at finite $E$ is dependent not only on the current $E$ but also on its applied history, yielding a hysteretic loop. From Fig. 3d and 3e, we find no obvious hysteresis within the sweep range $|E|_{max} < 28$ mV nm$^{-1}$. This feature is a typical ferroelectric characteristic that switching polarization necessitates $E$ exceeding a critical value. With increasing the sweep range of $|E|_{max} > 28$ mV nm$^{-1}$, the hysteresis loops and the corresponding $P_r$ dramatically increase, resulting in a large memory window with two non-volatile states. All the above features, including both spontaneous polarization and its switchable characteristic, unambiguously confirm the observations of ferroelectricity in our system.

When $|E|_{max}$ is increased to ~73 mV nm$^{-1}$, $P_r$ approaches the saturation polarization ($P_s$). Further increasing $|E|_{max}$ does not change $P_r$ anymore. However, the window of hysteresis loops continuously enlarges without showing any sign of saturation with increasing $|E|_{max}$. Figure 3e summarizes the $P_r$ and $P_s$ as a function of $|E|_{max}$, showing a step-like increase in $P_r$ and $P_s$. By further temperature-dependent measurements shown in Fig. 3f and Extended Data Fig. 5, $P_r$ decreases with increasing temperature at $|E|_{max} = 80$ mV nm$^{-1}$, which is opposite to other extrinsic mechanisms such as charge trapping and external adsorbates, but consistent with the typical ferroelectric behavior (see discussion in Methods). Moreover, the ferroelectric hysteresis loops and spontaneous polarization could survive even at room temperature, rendering our devices valuable in practical applications.

Besides the standard hysteresis loop, we also observed several unique switching behaviors. Firstly, conventional ferroelectric insulators always show an anticlockwise hysteretic $P - E$ loop because the bound charges only occur at the interface. However, ferroelectric semiconductors or metals, due to the presence of mobile carriers, can exhibit partial polarization switching, resulting in clockwise hysteresis [36]. In our system, the direction of hysteretic loops is always clockwise as shown in Fig. 3b and 3c, as a result of itinerant charges in graphene. The feature indicates the electric dipoles and itinerant carriers are in the same channel in our system. It's noteworthy that whether the hysteretic loops are clockwise or anticlockwise is highly relative to whether the specific gate is at the top or bottom. Secondly, above the threshold ($|E|_{max} > 73$ mV nm$^{-1}$), the $P_r$ is approximately identical to the $P_s$, indicating the existence of single domain in our sample. The uniform polarization and smooth switching in our system suggests that the ferroelectricity arises more likely from electronic dynamics, rather than sliding ferroelectricity assisted by domain motions. Thirdly, the hysteresis loops are highly dependent on $|E|_{max}$, which arises from the moiré trapping effects (see detailed mechanisms in Methods).

**Layer dependence of electronic ferroelectricity**

The GSAS and ferroelectricity observed in monolayer graphene moiré superlattice resemble that in bilayer counterpart [16,26]. Based on previous understanding, layer-polarized flat moiré bands and interlayer charge transfer are essential roles in the emergence of ferroelectricity in graphene moiré systems. To explicitly unveil layer-dependent ferroelectricity, we intentionally designed a control device, which allows us to *in-situ* compare the emerging ferroelectricity in monolayer, bilayer, and trilayer graphene moiré structures. It's found that all of these three systems exhibit ferroelectric behavior and have some characteristic in common (see Extended Data Fig. 7-10). Firstly, the ferroelectric hysteresis coexists with GSAS. Secondly, the GSAS depends on the gate sweep range and is believed to arise from the asymmetric moiré superlattice. Thirdly, the window of hysteresis loops continuously enlarges without saturation with the increasing $|E|_{max}$ as shown in Fig. 3d and Extended Data Fig. 7, which is an atypical behavior distinct from conventional ferroelectricity. Above a critical $|E|_{max}$, they have the same saturation Hall carrier density $n_H^s$, which is independent of layer number, but highly relative to half-filling density of a moiré band (see Methods). Our results suggest the electronic ferroelectricity observed here is independent of graphene layer number and the fine band structure of graphene (see discussion in Supplementary Information). However, we believe the semimetal characteristics of graphene plays a crucial role, since the formation of electron-hole dipoles hinges on the facile excitation of electron-hole pairs by gates.

**Memory device and synapse emulation**

As shown in Fig. 3e, $P_r$ can be continuously tuned from 0 to $P_s$ by changing $|E|_{max}$, distinct from two-state switching in conventional ferroelectric materials. The multiple spontaneous polarization states observed here are not arising from domain configurations as that in conventional ferroelectric memristors [37]. Instead, the tunable $P_r$ is achieved by the continuous injection of localized holes (electrons) by the specific gate ($V_t$, in this specific device) and electrons (holes) by the other gate (see discussion in Methods). This unique feature endows us to utilize it to realize unconventional applications. In the following, we demonstrate proof-of-concept devices, providing promising applications of monolayer graphene ferroelectricity in multi-state data storages and neuromorphic synapses.

We first establish the way to program $n_H$ with quasi-continuous multiple states by applying various pulses of $E$. As shown in Fig. 4b, positive and negative pulses of $E$ are sequentially applied with gradual increase in the amplitudes. Under pulses of small $E < 60$ mV nm$^{-1}$, $|n_H|$ increases quasi-continuously with increasing $|E|$. Each polarization state is stable over time even after $E$ is removed and switchable by changing sign of $E$ pulse shown in Fig. 4c. Such feature is a typical nonvolatile memory state, which can be widely applied to data storage. Beyond the simple two-state (0 and 1 in digital circuits) storage, our device can function as multiple state storage as demonstrated in Fig. 4a and 4b.

Our ferroelectric devices exhibit diverse electrical characteristics, enabling us to emulate multifunctional synaptic activities such as synaptic plasticity including LTP, LTD, and STP. By programming the magnitude and sign of $E$ pulses as shown in Fig. 4e, the non-volatile and quasi-continuous change of $|n_H|$ can emulate the synaptic plasticity, which is the key component of artificial synaptic devices for neuromorphic computing. In neuroscience, the long-term synaptic plasticity represents the ability of synapses to strengthen (LTP) or weaken (LTD) over time in response to increases or decreases in their activity and is widely considered as a primary mechanism for learning and memory. Analogously, in our system, by writing $E$ pulses, the device shows potentiation and depression of $|n_H|$ (emulating synaptic weight), which is reminiscent of LTP and LTD in synapse transistors, respectively.

More excitingly, we notice that $|n_H|$ experiences spontaneous decay in a short time at large $E > 60$ mV nm$^{-1}$, as shown in Fig. 4b. To better understand the behavior under large electric field, we investigated how $|n_H|$ emulating the synaptic weight evolves upon application of a train of $E$ pulses. Upon excitation of a high $E$ pulse, $|n_H|$ immediately increases to a high value and gradually relaxes until the application of next pulse as shown in Fig. 4f. This feature can be used to emulate STP for processing temporal information. This process is highly reproducible and can work for over 20 cycles without any degradation in performance as demonstrated in Fig. 4f. The underlying mechanism is still to be understood but may be attributed to the saturation of moiré traps.

**Outlook**

Our findings of electronic ferroelectricity in monolayer graphene not only enrich its fruitful properties in this wonder material, but also offers new opportunities for exploring novel physics. When interplaying with other properties in graphene such as ferromagnetism, topology, and superconductivity, intriguing properties including unconventional multiferroics and topological ferroelectrics may emerge. From the perspective of applications, monolayer graphene based on non-volatile memory and synaptic devices possess unique advantages in terms of high mobility, high stability and multifunctionality. Moreover, our graphene ferroelectric devices approach to 2D limit, compatible with post Moore's law era devices. As aforementioned, the main ferroelectric features in monolayer graphene moiré superlattices can survive up to room temperature, further facilitating their applications. The already well-developed growth of scalable monolayer graphene superlattices can be used to construct the novel ferroelectric devices [38]. The h-BN required for the construction of moiré superlattices also serves as the dielectric material, which simplifies the design of ferroelectric devices.

**References**

1. A. I. Khan, A. Keshavarzi & S. Datta. The future of ferroelectric field-effect transistor technology. *Nat. Electron.* **3**, 588-597, (2020).
2. D. Zhang, P. Schoenherr, P. Sharma & J. Seidel. Ferroelectric order in van der Waals layered materials. *Nat. Rev. Mater.* **8**, 25-40, (2023).
3. C. Wang, L. You, D. Cobden & J. Wang. Towards two-dimensional van der Waals ferroelectrics. *Nat. Mater.* **22**, 542-552, (2023).
4. Y. Cao *et al.* Nematicity and competing orders in superconducting magic-angle graphene. *Science* **372**, 264-271, (2021).
5. Y. Jiang *et al.* Charge order and broken rotational symmetry in magic-angle twisted bilayer graphene. *Nature* **573**, 91-95, (2019).
6. Y. Cao *et al.* Unconventional superconductivity in magic-angle graphene superlattices. *Nature* **556**, 43-50, (2018).
7. A. L. Sharpe *et al.* Emergent ferromagnetism near three-quarters filling in twisted bilayer

graphene. *Science* **365**, 605-608, (2019).
8. J. Valasek. Piezo-electric and allied phenomena in Rochelle salt. *Phys. Rev.* **17**, 475-481, (1921).
9. A. I. Khan *et al.* Negative capacitance in a ferroelectric capacitor. *Nat. Mater.* **14**, 182-186, (2015).
10. X. Yan *et al.* Moiré synaptic transistor with room-temperature neuromorphic functionality. *Nature* **624**, 551-556, (2023).
11. K. Chang *et al.* Discovery of robust in-plane ferroelectricity in atomic-thick SnTe. *Science* **353**, 274-278, (2016).
12. F. Liu *et al.* Room-temperature ferroelectricity in $CuInP_2S_6$ ultrathin flakes. *Nat. Commun.* **7**, 12357, (2016).
13. Z. Fei *et al.* Ferroelectric switching of a two-dimensional metal. *Nature* **560**, 336-339, (2018).
14. S. Yuan *et al.* Room-temperature ferroelectricity in $MoTe_2$ down to the atomic monolayer limit. *Nat. Commun.* **10**, 1775, (2019).
15. J. Gou *et al.* Two-dimensional ferroelectricity in a single-element bismuth monolayer. *Nature* **617**, 67-72, (2023).
16. Z. Zheng *et al.* Unconventional ferroelectricity in moire heterostructures. *Nature* **588**, 71-76, (2020).
17. K. Yasuda, X. Wang, K. Watanabe, T. Taniguchi & P. Jarillo-Herrero. Stacking-engineered ferroelectricity in bilayer boron nitride. *Science* **372**, 1458-1462, (2021).
18. Y. W. M. Vizner Stern, W. Cao,I.Nevo, K. Watanabe, T. Taniguchi, E. Sela, M. Urbakh,O. Hod, M. Ben Shalom. Interfacial ferroelectricity by van der Waals sliding. *Science* **372**, 1462-1466, (2021).
19. A. Weston *et al.* Interfacial ferroelectricity in marginally twisted 2D semiconductors. *Nat. Nanotechnol.* **17**, 390-395, (2022).
20. X. Wang *et al.* Interfacial ferroelectricity in rhombohedral-stacked bilayer transition metal dichalcogenides. *Nat. Nanotechnol.* **17**, 367-371, (2022).
21. L. Rogée *et al.* Ferroelectricity in untwisted heterobilayers of transition metal dichalcogenides. *Science* **376**, 973-978, (2022).
22. C. R. Woods *et al.* Charge-polarized interfacial superlattices in marginally twisted hexagonal boron nitride. *Nat. Commun.* **12**, 347, (2021).
23. T. H. Yang *et al.* Ferroelectric transistors based on shear-transformation-mediated rhombohedral-stacked molybdenum disulfide. *Nat. Electron.* **7**, 29-38, (2024).
24. A. Jindal *et al.* Coupled ferroelectricity and superconductivity in bilayer $Td-MoTe_2$. *Nature* **613**, 48-52, (2023).
25. R. Niu *et al.* Giant ferroelectric polarization in a bilayer graphene heterostructure. *Nat. Commun.* **13**, 6241, (2022).
26. Z. Zheng *et al.* Electronic ratchet effect in a moiré system: signatures of excitonic ferroelectricity. *arXiv*, arXiv:2306.03922, (2023).
27. P. Wang *et al.* Moiré synaptic transistor for homogeneous-architecture reservoir computing. *Chin. Phys. Lett.* **40**, 117201, (2023).
28. D. R. Klein *et al.* Electrical switching of a bistable moiré superconductor. *Nat. Nanotechnol.* **18**, 331-335, (2023).
29. Y. Wang *et al.* Ferroelectricity in hBN intercalated double-layer graphene. *Front. Phys.* **17**,

43504, (2022).

30. Z. Zhu, S. Carr, Q. Ma & E. Kaxiras. Electric field tunable layer polarization in graphene/boron-nitride twisted quadrilayer superlattices. *Phys. Rev. B* **106**, 205134, (2022).
31. B. Hunt *et al.* Massive Dirac fermions and Hofstadter butterfly in a van der Waals heterostructure. *Science* **340**, 1427-1430, (2013).
32. L. Wang *et al.* Evidence for a fractional fractal quantum Hall effect in graphene superlattices. *Science* **350**, 1231-1234, (2015).
33. L. A. Ponomarenko *et al.* Cloning of Dirac fermions in graphene superlattices. *Nature* **497**, 594-597, (2013).
34. C. R. Dean *et al.* Hofstadter's butterfly and the fractal quantum Hall effect in moiré superlattices. *Nature* **497**, 598-602, (2013).
35. A. I. Berdyugin *et al.* Out-of-equilibrium criticalities in graphene superlattices. *Science* **375**, 430-433, (2022).
36. M. Si *et al.* A ferroelectric semiconductor field-effect transistor. *Nat. Electron.* **2**, 580-586, (2019).
37. A. Chanthbouala *et al.* A ferroelectric memristor. *Nat. Mater.* **11**, 860-864, (2012).
38. W. Yang *et al.* Epitaxial growth of single-domain graphene on hexagonal boron nitride. *Nat. Mater.* **12**, 792-797, (2013).
39. M. Si, X. Lyu & P. D. Ye. Ferroelectric Polarization Switching of Hafnium Zirconium Oxide in a Ferroelectric/Dielectric Stack. *ACS Appl. Electron. Mater.* **1**, 745-751, (2019).
40. P. Rickhaus *et al.* The electronic thickness of graphene. *Sci. Adv.* **6**, eaay8409, (2020).
41. A. C. Ferrari & D. M. Basko. Raman spectroscopy as a versatile tool for studying the properties of graphene. *Nat. Nanotechnol.* **8**, 235-246, (2013).
42. N. R. Finney *et al.* Tunable crystal symmetry in graphene-boron nitride heterostructures with coexisting moire superlattices. *Nat. Nanotechnol.* **14**, 1029-1034, (2019).
43. H. Wang, Y. Wu, C. Cong, J. Shang & T. Yu. Hysteresis of Electronic Transport in Graphene Transistors. *ACS Nano* **4**, 7221-7228, (2010).

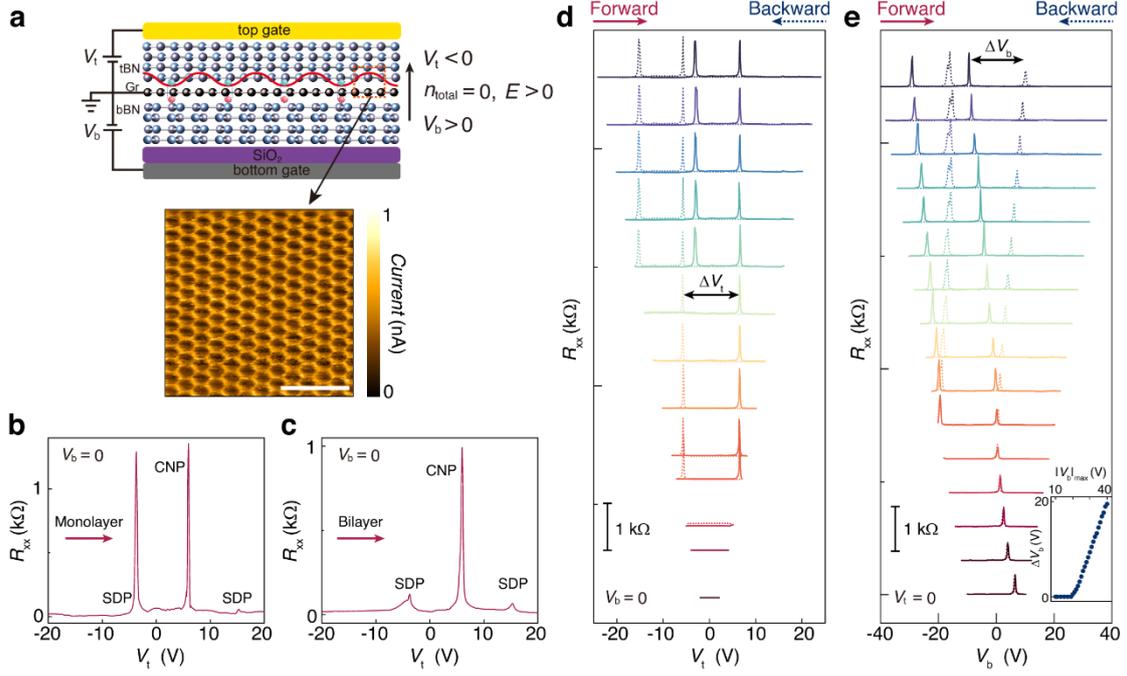

**Fig. 1 | Ferroelectric hysteresis in monolayer graphene superlattices. a**, Schematic of our device with asymmetric moiré interfaces and dual-gate structure. The red wavy line illustrates the moiré potential at top interface trapping holes (blue balls) injected by the top gate ($V_t$), which are bound by itinerant electrons (red balls) injected by the bottom gate ($V_b$). The vertical arrow defines the positive electric field. Bottom panel shows conductive atomic force microscopy image of graphene/h-BN moiré superlattices. The scale bar is 50 nm. **b,c**, Four-terminal longitudinal resistance $R_{xx}$ as a function of $V_t$ at fixed $V_b = 0$ for monolayer graphene (**b**) and bilayer graphene (**c**) moiré superlattices. The arrows illustrate the sweep direction. **d**, $R_{xx}$ as a function of $V_t$ by sweeping $V_t$ in various ranges ($|V_t|_{max}$) while fixing $V_b = 0$. **e**, $R_{xx}$ as a function of $V_b$ by sweeping $V_b$ in various ranges ($|V_b|_{max}$) while fixing $V_t = 0$. The curves in (**d**) and (**e**) are vertically shifted for clarity. The forward and backward sweeps are shown in solid and dashed lines, respectively. The inset in (**e**) plots the difference of charge-neutrality points between forward and backward sweeps as a function of $|V_b|_{max}$.

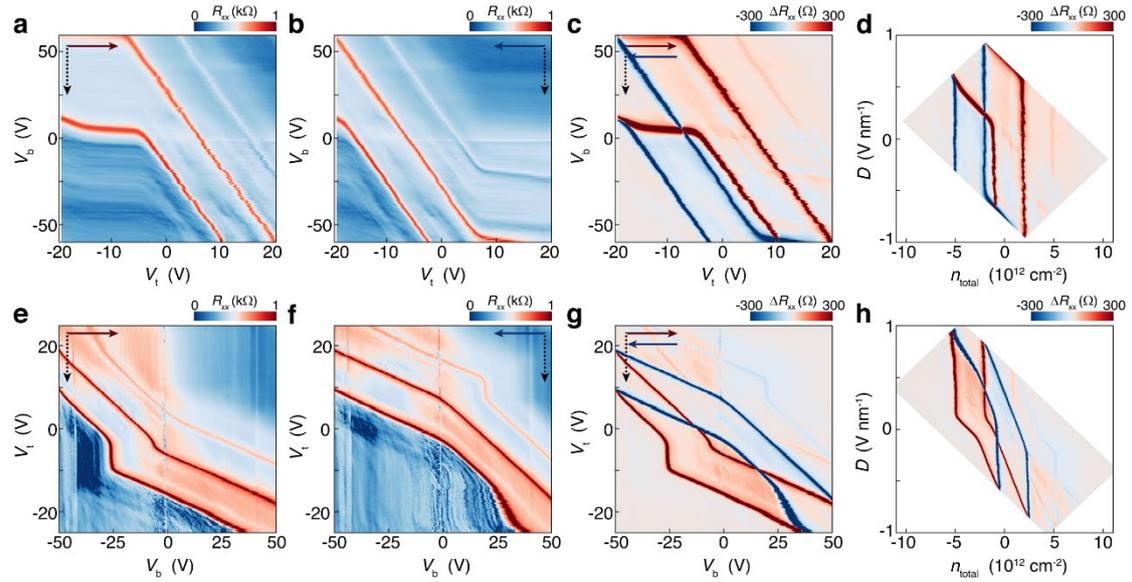

**Fig. 2 | Gate-specific anomalous screening and scan-dependent hysteresis. a,b**, Dual-gate maps of $R_{xx}$ by scanning $V_t$ forward (**a**) and backward (**b**) at each fixed $V_b$. **c**, The difference between $R_{xx}$ in (**a**) and (**b**). **d**, The corresponding $n_{\text{total}} - D$ plot of (**c**). **e,f**, Dual-gate maps of $R_{xx}$ by scanning $V_b$ forward (**e**) and backward (**f**) at each fixed $V_t$. **g**, The difference between $R_{xx}$ in (**e**) and (**f**). **h**, The corresponding $n_{\text{total}} - D$ plot of (**g**). In (**a**)-(**c**) and (**e**)-(**g**), the fast-scan axis and slow-scan axis are plotted in horizontal axis and vertical axis, respectively. The solid arrows illustrate the fast-scan direction, and the dashed arrows marks the slow-scan direction.

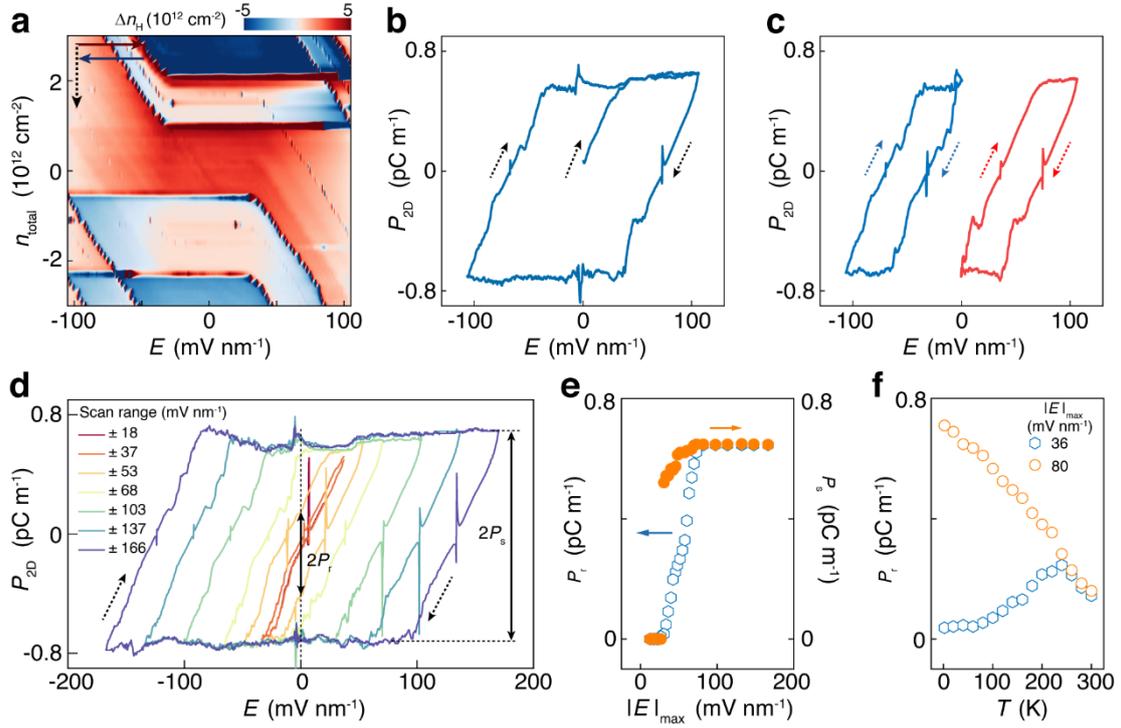

**Fig. 3 | Polarization-electric field ($P-E$) hysteresis loops. a**, The difference of Hall carrier density $n_H$ between the forward and backward sweeps of external electric field $E$ at each fixed carrier density $n_{total}$. $n_H = -\frac{B}{eR_{xy}}$ is obtained by measuring anti-symmetrized Hall resistance $R_{xy}$ at $B = \pm 0.1$ T. **b**, Two-dimensional polarization $P_{2D}$ as a function of $E$ measured by sweeping $E$ sequentially in the direction denoted by the arrows at fixed $n_{total} = 0$. $P_{2D} = en_H d_{dipole}$ is obtained by measuring $n_H$, where $d_{dipole}$ is the thickness of monolayer graphene. **c**, Similar data to (**b**) but for initial polarization at positive (blue curve) and negative (red curve) values. The sweeping sequence is marked by the dashed arrows. **d**, Scan-range dependent $P-E$ hysteresis loops measured by the same method as that in (**b**). The remanent polarization $P_r$ and saturation polarization $P_s$ are extracted according to the marks in the plots. **e**, $P_r$ and $P_s$ as a function of scan range $|E|_{max}$. No hysteresis is observed at $|E|_{max} < 28$ mV nm$^{-1}$, resulting in zero $P_r$ and $P_s$. Above $|E|_{max} > 73$ mV nm$^{-1}$, $P_r$ approaches $P_s$, indicating the single ferroelectric domain. **f**, Temperature dependent $P_r$ at two representatives $|E|_{max} = 36$ mV nm$^{-1}$ and 80 mV nm$^{-1}$. $P_r$ can be nonzero even at room temperature, indicating the room-temperature ferroelectricity in our system. Data in (**e**) and (**f**) were acquired at different cooling down cycles. All the data in (**b**)-(**f**) were measured at fixed $n_{total} = 0$. The sharp peaks near $P_{2D} = 0$ in (**b**)-(**d**) are due to the measured $R_{xy} \to 0$ when passing through CNP.

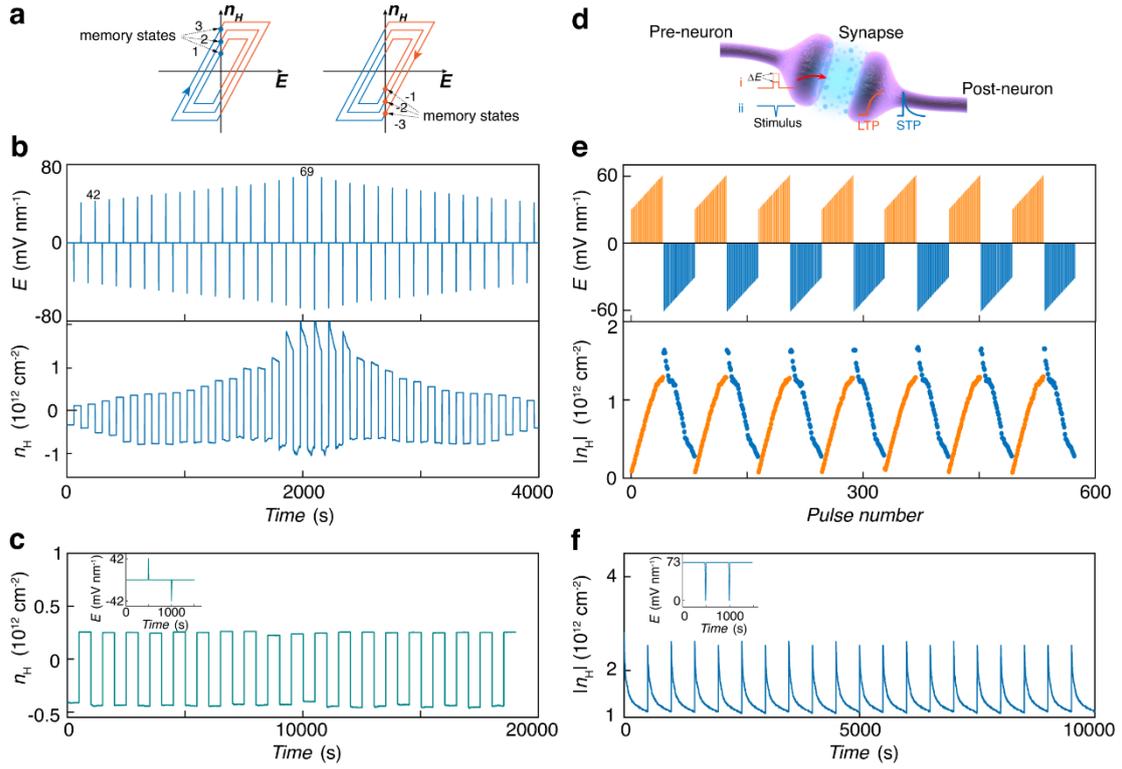

**Fig. 4 | Multi-state data storage and multifunctional synaptic devices. a**, Schematic of multi-state remanent polarizations. **b**, Programmable measurement of $n_H$ in response to a series of $E$ pulses. $E$ pulses were programmed to be alternate positive and negative values with gradually increasing amplitudes. The bottom panel is the corresponding $n_H$ as a function of time. Within small $E$ range, $n_H$ shows nonzero value even after $E$ returning to zero, namely, exhibiting non-volatile characteristic. The amplitude of $n_H$ gradually increases with the increasing amplitude of $E$, meanwhile its sign is changed with reversing $E$. Working in this regime, the device can be used as multi-state data storage. At $|E| > 60$ mV nm$^{-1}$, the corresponding $n_H$ decays with time. This regime can be used as the emulation of short-term plasticity in synaptic devices. **c**, $n_H$ measured as a function of time under repetitive $E$ pulses alternating between two equal values in amplitude but opposite sign. The inset shows one cycle of applied $E$. **d**, Schematic of a biological synapse. **e**, The emulation of long-term plasticity in synaptic devices by utilizing non-volatile multiple states in our device. $|n_H|$ serves as the synaptic weights, which can continuously increase and decrease under a series of input stimuli (here is $E$), emulating the long-term potentiation and long-term depression, respectively. **f**, The emulation of short-term plasticity. $|n_H|$ evolves as a function of time under repetitive stimuli of $E$. The inset shows one cycle of applied $E$. All the $n_H$ in (**b**), (**c**), (**e**) and (**f**) are measured from $R_{xy}$ at $B = 0.1$ T without anti-symmetric processing. Therefore, there is slightly difference between positive and negative $n_H$ due to the unperfect Hall bar geometry. All the data are measured at fixed $n_{\text{total}} = 0$.

**Methods:**

**Device fabrication**

All the devices were made using standard dry-transfer method. In brief, graphene and h-BN flakes were mechanically exfoliated on the 285 nm SiO$_2$/Si substrates. In particular, the graphene flake of Device D1 contains consecutive mono-/bi-/tri-layer parts, which were identified by their optical contrast (Extended Data Fig. 1). The top h-BN, graphene, and bottom h-BN flakes were layer-by-layer assembled using a poly(bisphenol A carbonate)/polydimethylsiloxane (PC/PDMS) stamp. Graphene flake was intentionally aligned with both top and bottom h-BN by utilizing their straight and long edges. The alignment between graphene and h-BN can be determined by transport measurement, while the relative angle between top and bottom h-BN were determined by second harmonic generation (SHG) measurements. The h-BN/graphene/h-BN heterostructure was finally released on a plasma-cleaned SiO$_2$/Si substrate which served as back gates. The contact regions were patterned by electron beam lithography (EBL) and etched as trenches by CHF$_3$/O$_2$ plasma. Metallic contacts (5 nm Cr/ 60 nm Au) were deposited into the trenches. The metallic top gate was made by a second EBL and e-beam evaporation. The final Hall bar geometry was defined by another round of EBL and plasma etching. The list of devices, including the ferroelectricity behaviors and the stagger angles between top and bottom h-BN, are summarized in Table S1.

**Optical measurements**

Raman measurements were performed at room temperature using a confocal Raman spectrometer (Witec Alpha 300RAS with UHTS 300 spectrometer) equipped with a 532 nm excitation laser. The laser spot size was about 1 μm. The second harmonic generation (SHG) measurements were conducted in the same setup with incident light wavelength of 1064 nm and a fixed excitation power of 20 mW.

**Conductive atomic force microscopy measurements:**

The sample for conductive atomic force microscopy (c-AFM) measurements was made by sequentially picking up h-BN and graphene with PC/PDMS stamp. The stack was then flipped on a fresh PDMS. After dissolving PC with n-methyl-2-pyrrolidone (NMP), the stack was released on a new SiO$_2$/Si substrate. The c-AFM measurements were performed on Asylum Research Jupiter XR at room temperature.

**Transport measurements**

The devices were wire-bonded on the LCC-44 chip carriers (Kyocera) and loaded on a closed-cycle cryostat. The longitudinal and Hall resistances were measured by lock-in amplifier (SR-830) with constant excitation current of 100 nA. The top and bottom gates were applied by source meters (Keithley 2450).

Our dual-gate structure allows us to convert $V_t - V_b$ maps to $n_{\text{total}} - D$ maps. The total gate-induced carrier density is $n_{\text{total}} = (C_b V_b + C_t V_t)/e$, and the displacement field is $D = (C_b V_b - C_t V_t)/2\varepsilon_0$, where $C_t$ and $C_b$ are the capacitance per area of top and bottom gate measured by Hall effect, respectively, $e$ is the elementary charge, and $\varepsilon_0$ is the vacuum permittivity. The four-terminal longitudinal resistance $R_{xx}$ maps were acquired at zero magnetic field, unless otherwise specified.

To convert $D$ to $E$, we note that $V_b$ was applied on SiO$_2$ and bottom h-BN. Therefore, the voltage across the bottom h-BN is $V_{bBN} = \frac{V_b \varepsilon_{SiO_2} d_{bBN}}{\varepsilon_{BN} d_{SiO_2} + \varepsilon_{SiO_2} d_{bBN}}$, where $d_{bBN}$ and $d_{SiO_2}$ are the thickness of bottom h-BN and SiO$_2$, respectively [39], $\varepsilon_{SiO_2} = 4.3$ and $\varepsilon_{BN} = 3.8$ are the dielectric constant of h-BN and SiO$_2$, respectively, calibrated by capacitance measurements through Hall effect at normal gating region. The electric field across graphene is $E = (V_{bBN}/d_{bBN} - V_t/d_{tBN})/2 = D/\varepsilon_{BN}$, where $d_{tBN}$ is the thickness of top h-BN.

### $n_H$ and $P_{2D} - E$ loops measurements

The itinerant carrier density $n_H$ is determined from Hall effect by measuring Hall resistance $R_{xy}$ under small magnetic field of $B = \pm 0.1$ T. $R_{xy}$ is anti-symmetrized by $R_{xy} = (R_{xy}^{0.1T} - R_{xy}^{-0.1T})/2$ to remove residual values induced by unperfect sample geometry. The Hall (itinerant) carrier density is calculated to be $n_H = -B/eR_{xy}$. To measure Fig. 3a, we simultaneously swept $V_b$ and $V_t$ with a specific relation by fixing $n_{total} = (C_b V_b + C_t V_t)/e$ as a constant value, and measured $n_H$. In this way, we can realize the sweeps of $E$ at fixed $n_{total}$, namely, $E$ as fast-scan axis and $n_{total}$ as slow-scan axis. The $\Delta n_H$ is the difference of $n_H$ between forward and backward sweeps of $E$.

In order to measure $P_{2D} - E$ loops, we always fixed $n_{total} = 0$. By sweeping $E$ forward and backward in a given rang of $|E|_{max}$, we acquired the response of $n_H$. The 2D polarization $P_{2D}$ follows $P_{2D} = e n_H d_{dipole}$ [17,20], where $d_{dipole}$ is the size of dipole moment, i.e., the thickness of graphene. For monolayer graphene, $d_{dipole} = 0.26$ nm [40], while for bilayer graphene, $d_{dipole}$ is doubled as that of monolayer graphene.

### Distinguishing monolayer from bilayer graphene

Although unconventional ferroelectricity has been observed in Bernal bilayer graphene/h-BN superlattice, its underlying mechanism is still mysterious. Previous understanding is that layer-polarized flat moiré bands and interlayer charge transfer between top and bottom graphene layers play an important role [16,26]. Therefore, bilayer graphene is the key factor. One of our main findings is that monolayer graphene exhibits similar ferroelectricity, thus breaking previous understanding. Therefore, it's important to ensure what we measured was indeed monolayer graphene transport behavior. We used several features to distinguish monolayer and bilayer graphene.

Firstly, the layer number of graphene can be determined via reflection contrast spectroscopy. As depicted in Extended Data Fig. 1b and 1c, the exfoliated graphene flake shows consecutive steps. The optical contrast relative to the 285 nm SiO$_2$/Si substrate decreases with the increasing graphene layer number monotonously, following the Beer-Lambert law. The layer-number-dependent optical contrast helps us distinguish the monolayer from the bilayer and tri-layer graphene. Our transport measurements from monolayer, bilayer, and trilayer graphene were performed from the devices sharing the same stack, which allows us to *in-situ* compare their characteristics.

Secondly, Raman spectra are capable of determining layer number of graphene for layer thickness of less than four layers [41]. The 2D peaks of graphene in Raman spectra are sensitive to the layer number. As shown in Extended Data Fig. 1f, the 2D peak from monolayer graphene is sharper and exhibits slightly redshift, compared with bilayer graphene. Moreover, it can be fitted by a single Lorentz function with a full width at half maximum (FWHM) of 32 cm$^{-1}$, which is a typical feature of single-aligned monolayer graphene/h-BN stack [42].

Thirdly, the transport behaviors of monolayer graphene and bilayer graphene moiré superlattice exhibit distinguishable features. Generally, the peak resistance of SDP at hole side is comparable with that of CNP and SDP shows prominent electron-hole asymmetry in monolayer graphene/h-BN superlattice [31-33,35]. By contrast, bilayer graphene/h-BN superlattice exhibits weak hole-side SDP [34]. This gives us an empirical method of distinguishing monolayer and bilayer graphene directly from transport measurements (see Fig. 1b-c).

Fourthly, in bilayer graphene, the external electric field can open a gap at CNP with its size increases with increasing electric field. By contrast, the CNP of monolayer graphene is almost resilience to the electric field. We have measured the dual-gate dependent $R_{xx}$ near CNP for monolayer and bilayer graphene, shown in Extended Data Fig. 2. As expected, the CNP peak increases dramatically with electric field in bilayer graphene, while remaining in the same order of magnitude in monolayer counterpart.

Overall, we can unambiguously confirm that the measurements in the main text are from monolayer graphene/h-BN. Therefore, we conclude the unconversional ferroelectricity can emerge in monolayer graphene/h-BN superlattice.

**Crystallographical Alignment and twist angle determination**

We utilized Raman spectra and SHG to figure out the alignment between graphene and h-BN. During the van der Waals assembly process, we intentionally aligned graphene with both top and bottom h-BN by using their straight edges. The result can be either single alignment or doube alignement. The Raman spectra of monolayer part in final stack yields a 2D peak with FWHM of 32 cm$^{-1}$ (Extended Data Fig. 1f), indicating the single alignment [42]. We further confirmed the top h-BN and bottom h-BN have a relative angle of 30° by SHG measurements as shown in Extended Data Fig. 1d and 1e.

To determine whether the top h-BN or bottom h-BN is aligned with graphene, we intentionally reserved one part of monolayer graphene covered solely by top h-BN without any contact with bottom h-BN, as shown in Extended Data Fig. 1a. Our Raman spectra measured from this part shows a 2D peak with FWHM of 32 cm-1 (Extended Data Fig. 1g), which is a typical feature of graphene/h-BN superlattice [42]. For reference, typcial non-aligned graphene has a 2D peak with FWHM of 20 cm$^{-1}$ [42]. With this, we can confirm that graphene was singly aligned with the top h-BN for Device D1.

To precisely determine the twist angle between graphene and top h-BN, we directly resort to the transport results. The moiré wavelength of moiré superlattice between graphene and h-BN is determined by $\lambda = \frac{(1+\delta)a_G}{\sqrt{2(1+\delta)(1-\cos\theta)+\delta^2}}$, where $a_G = 0.246$ nm is the in-plane lattice constant of graphite, $\delta \approx 1.6\%$ is the lattice mismatch between graphene and h-BN, and $\theta$ is the relative misalignment angle between the two lattices. From the transport measurement, we found the full filling carrier density of moiré superlattice in Device D1 was $n_{\text{Full}} = 3.0 \times 10^{12}$ cm$^{-2}$, determined from the difference between hole-side SDP and CNP at normal regime without screening effect as shown in Fig. 2d and 2h. The area of a moiré unit cell is $A = 4/n_{\text{Full}}$. Then we can obtain the moiré wavelength is 12.4 nm, by using $A = \frac{\sqrt{3}}{2}\lambda^2$. The twist angle between graphene and top h-BN is determined to be 0.68°.

**Excluding extrinsic effects that account for the hysteresis**

Gate hysteresis arising from extrinsic effects, such as charge traps induced by defects, impurities or absorbates, strongly depends on the scan rates and scan cycles of the gate voltages [43]. In our devices, we observed robust gate hysteresis independent of scan rates and cycles. In Extended Data Fig. 3, we fixed $V_t = 0$ V and swept $V_b$ within the same scan range, but varying scan rates from 150 mV s$^{-1}$ to 400 mV s$^{-1}$. Obviously, individual forward and backward scan curves exhibit almost independence of scan rates in terms of both the resistance peak position and magnitude. In Extended Data Fig. 4, we also examine the endurance of the hysteresis by repetitively scanning $V_b$ or $V_t$ forward and backward for 20 cycles. All the curves overlap with each other, demonstrating their good cycle endurance. We observed similar robust hysteresis and good endurance in monolayer, bilayer and tri-layer graphene moiré superlattice, indicating they share the common origins.

In general, gate hysteresis induced by extrinsic effects should exhibit an enhanced performance at high temperature because the charges acquire high energy to migrate. The temperature-dependent hysteresis loops and GSAS in our systems exhibit clear weaker effects with increasing temperature, as shown in Extended Data Fig. 5 and Fig. S5 in Supplementary Information.

All the above features manifest that the hysteresis loops observed in graphene moiré superlattices are the intrinsic behavior stemming from polarization switching, rather than extrinsic effects.

**Mechanism of electronic ferroelectricity**

The measurements of $P_{2D} - E$ hysteresis loops provide us fruitful information to understand the mechanism of the observed ferroelectricity in our system. To this end, we re-plot Fig. 3b by using $n_H$ instead of $P_{2D}$, shown in Extended Data Fig. 6. We first discuss the polarization and switching process based on electronic ferroelectricity and then provide the corresponding evidence.

The hystersis loops can be devided into eight processes in a half cycle as shown in Extended Data Fig. 6a. The other half cycle has similar processes. We note that $n_H$ measured from Hall effect are the itinerant charges, mainly injected by $V_b$. In the whole processes, we fixed $n_{total} = 0$, namely, we attempt to inject equal density of holes (electrons) and electrons (holes) by $V_t$ and $V_b$, respectively. In normal dual-gating effect, we always have $n_H = n_{total} = 0$. However, since the moiré potential at the top interface localizes the carriers injected by $V_t$, the measured $n_H$ can be nonzero.

In Process i, localized holes and itinerant electrons are continuously injected by $V_t$ and $V_b$, respectively. $n_H$ is gradually increased, resulting in a nonzero polarization. The electron-hole dipole moments are formed by the localized holes at top interface and itinerant electrons at bottom interface. It's worth noting that the bound of localized hole and itinerant electron pairs is dynamic, resembling the formation of Cooper pairs in superconductivity.

When the density of the injected holes by $V_t$ approaches the half filling of the moiré band, Process ii is reached. Further increasing $E$ will trigger Process iii. In this process, the saturated moiré band cannot trap the additionally injected holes by $V_t$ any more. Instead, they recombine with the electrons injected by $V_b$. Therefore, $n_H$ is saturated, and the dual-gating effect becomes valid.

In Process iv, $E$ is decreased, namely, holes and electrons are extracted by $V_t$ and $V_b$, respectively.

The remaining itinerant electrons (i.e., $n_H$) continuously decrease till zero, arriving at Process v. Further decreasing $E$ will inject localized electrons at top interface and itinerant holes at the bottom interface, as shown in Process vi. The direction of electron-hole dipole moment (i.e. polarization) is reversed, relative to Process i-iv. In the subsequent Process vii-viii, $n_H$ saturates again, similar to Process ii-iii but with opposite values.

We found several experimental evidence which can support above mechanism.

(1) As shown in Fig. 3b and 3e, we always observed clockwise hysteresis loops in our device, different from the anticlockwise hysteresis loops in conventioanl ferroelectric insulators and sliding ferroelectric insulators [17,20]. Clockwise hysteresis loops can exist in ferroelectric semiconductors or metals, due to the presence of mobile carriers and partial polarization switching [36]. This feature indicates the electric dipoles and itinerant carriers are in the same channel in our system.

(2) The measured saturation itinerant density $n_H^s \approx 1.5 \times 10^{12}$ cm$^{-2}$ is independent of $|E|_{max}$ (see Fig. 3d) and layer number (see Extended Data Fig. 7), but highly relative to moiré period. In our device, the observation of SDP facilitates the determination of the full filling carrier density ($n_{Full} = 3.0 \times 10^{12}$ cm$^{-2}$) of a moiré band. We found that the saturation occurs at half filling of the moiré band, as we have $n_H^s \approx 0.5 n_{Full}$. Increasing temperature will thermally active the localized carriers, causing them to become itinerant ones. Therefore, we observed the decrease in $n_H^s$ (or $P_s$) with increasing temperature, as shown in Extended Data Fig. 5. The temperature dependent $n_H^s$ follows the thermal activation fitting as shown in Fig. S4 in Supplementary Inforamtion.

(3) From Extended Data Fig. 6a, the equality relation of $n_H^r = n_H^s$, indicates the uniform ferroelectric domain in our system. The slope of the curve tracing from Process iv-v-vi-vii remains unchanged, regardless of variations in $|E|_{max}$ (see Extended Data Fig. 6c), temperature (see Extended Data Fig. 5), or layer number (see Extended Data Fig. 7), as it's only relative to the gating capacity of $V_b$. The polarization switching is not due to the domain motion as that in conventional ferroelectricity or sliding ferroelectricity, but arising from the process of injecting and exctracting localized carriers assisted by GSAS as shown in Process iv-v-vi-vii.

(4) When $|E|_{max} < 73$ mV nm$^{-1}$, Process ii is absent, because of the insufficient supply of localized carriers in Process i, resulting in $n_H^r < n_H^s$. This is also the origin of the nonmonotonic dependence of $P_r$ on temperature for $|E|_{max} = 36$ mV nm$^{-1}$ as shown in Fig. 3f and Extended Data Fig. 5b, because it needs to keep the slope of the GSAS curves unchanged while decreasing $n_H^s$ with increasing temperature. Within the region of $n_H^r < n_H^s$, quasi-continuous remanent polarizations, which are stable even after $E$ is removed, endows the device with the functionality of mult-state data storage and mutlifunctional synapse emulation.

**Acknowledgements:**
This work was funded by National Natural Science Foundation of China (Grant No. 12274354), the Zhejiang Provincial Natural Science Foundation of China (Grant No. LR24A040003; XHD23A2001), and Westlake Education Foundation at Westlake University. We thank Chao Zhang from the Instrumentation and Service Center for Physical Sciences (ISCPS) at Westlake University for technical support in data acquisition. We also thank Westlake Center for Micro/Nano Fabrication and the Instrumentation and Service Centers for Molecular Science for facility support. K.W. and T.T. acknowledge support from the JSPS KAKENHI (Grant Numbers 21H05233 and 23H02052) and World Premier International Research Center Initiative (WPI), MEXT, Japan.

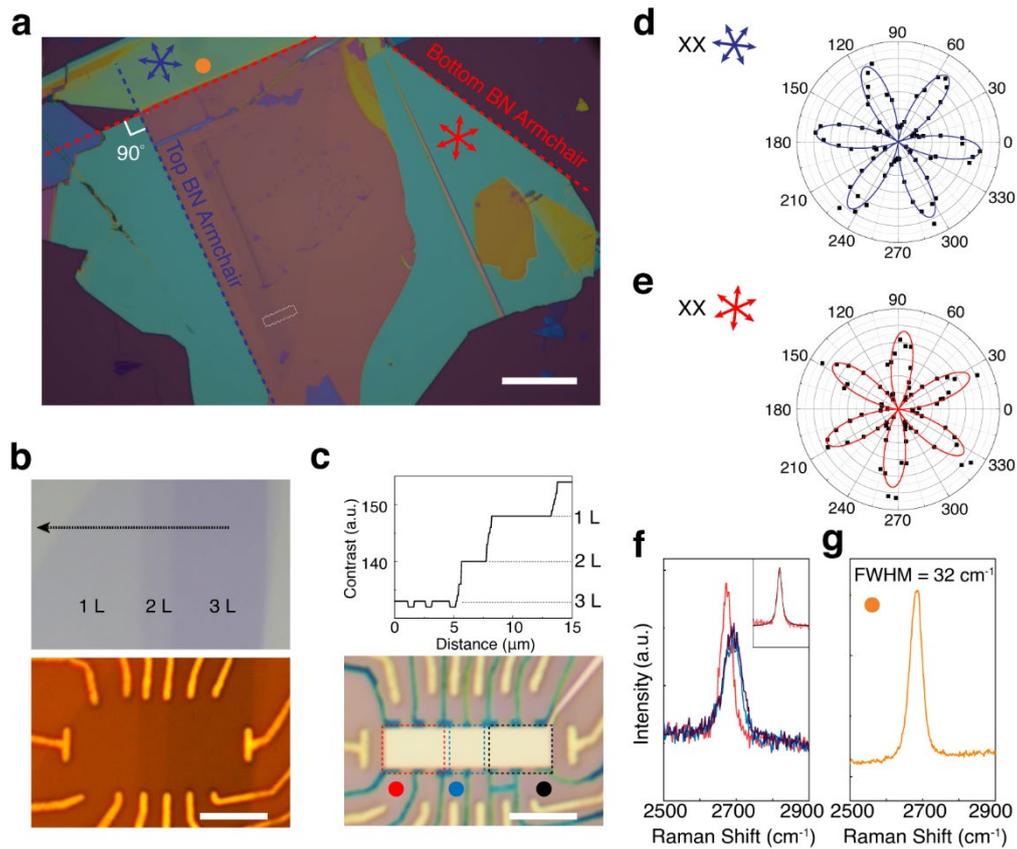

**Extended Data Fig. 1 | Optical characterizations of Device D1. a**, Optical image of the h-BN/graphene/h-BN heterostructure. The bule (red) dashed lines correspond to the armchair crystallographic axes of top (bottom) h-BN. The white dashed lines indicate the Hall bar configuration. The scale bar is 10 μm. **b**, Optical image of graphene flake before encapsulation and after nanofabrication, in which the mono-/bi-/tri-layer are labelled. The scale bar is 5 μm. **c**, Cross-sectional profile of optical contrast along the black dashed arrow in (**b**). Bottom panel: optical image of final Hall bar device, where the red/bule/black regions are mono-/bi-/tri-layer graphene channels, respectively. The scale bar is 5 μm. **d,e**, SHG signals for top and bottom h-BN, respectively. The relative angle between them is ~30°. **f**, Comparison of Raman spectra acquired in mono-/bi-/tri-layer graphene, marked as red/bule/black spots in (**c**). The inset shows the single-peak Lorentz fitting of Raman signal in monolayer graphene. **g**, Raman spectrum of monolayer graphene, yielding a full width at half maxima (FWHM) value of 32 cm$^{-1}$. The data were acquired at the orange spot in (**a**), where it's solely covered by top h-BN. With this, we can identify that graphene is singly aligned with top h-BN.

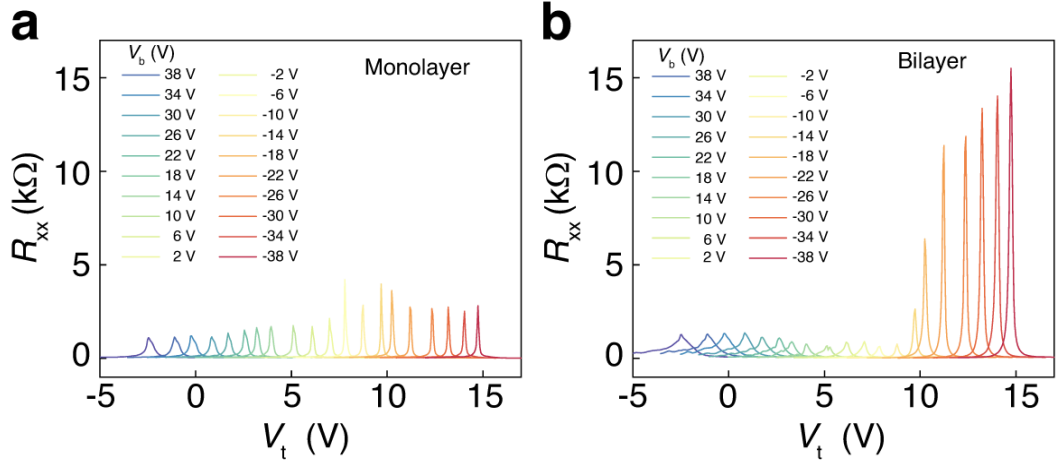

**Extended Data Fig. 2 | Field-dependent longitudinal resistance near CNP of mono- and bilayer graphene. a,b,** Longitudinal resistance of monolayer (**a**) and bilayer (**b**) graphene as a function of $V_t$ at fixed $V_b$ ranging from -38 V to 38 V. Data were collected by fixing $V_b$ and scanning $V_t$ from -15 V to 15 V. For better comparison, we only plotted the resistance peaks near CNP and kept them in the same plotting scale. Due to the screening effect, the resistance peaks of bilayer graphene are almost the same when $V_b$ decrease from 38 V to 2 V. When $V_b$ decrease further from -2 V to -38 V, the resistance peaks increase dramatically. In monolayer graphene, the resistance peaks of CNP remain in the same order of magnitude within the whole scanning range.

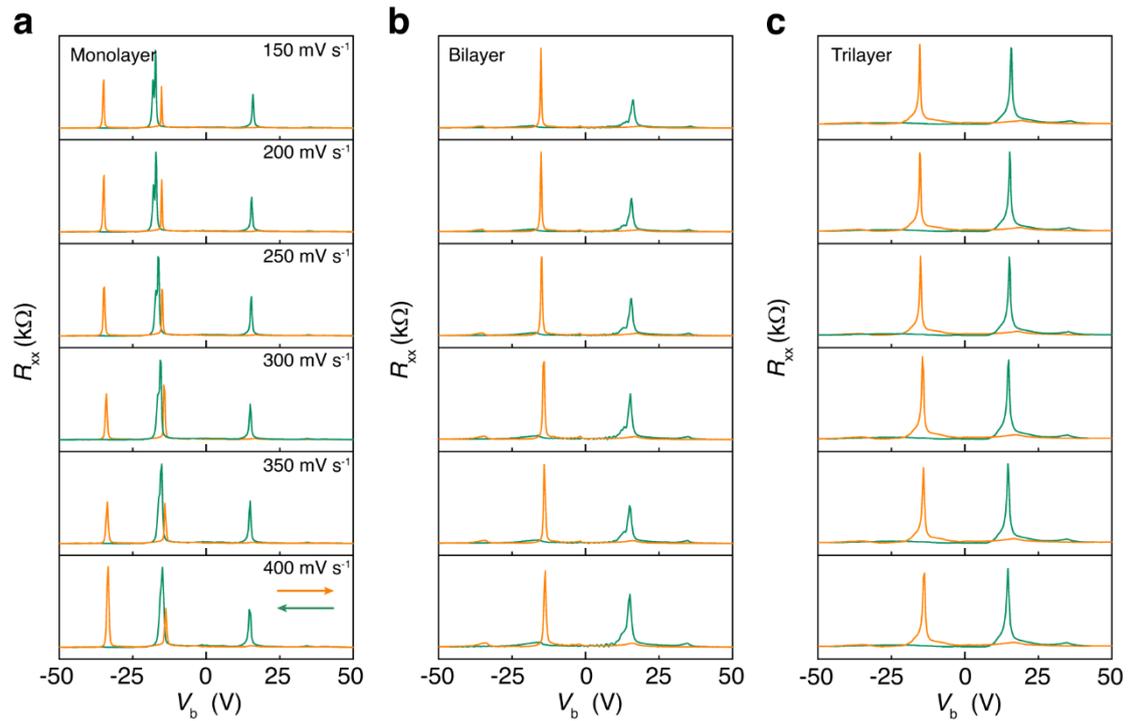

**Extended Data Fig. 3 | Independence of hysteresis on the scan rates of $V_b$. a-c**, Forward (orange) and backward (green) scans of $V_b$ for different scan rates in monolayer (**a**), bilayer (**b**) and tri-layer (**c**) graphene at a fixed $V_t = 0$ V. From upper panel to bottom panel, no noticeable variation was observed in terms of the position and magnitude of resistance peaks with the scan rates increasing from 150 mV s$^{-1}$ to 400 mV s$^{-1}$.

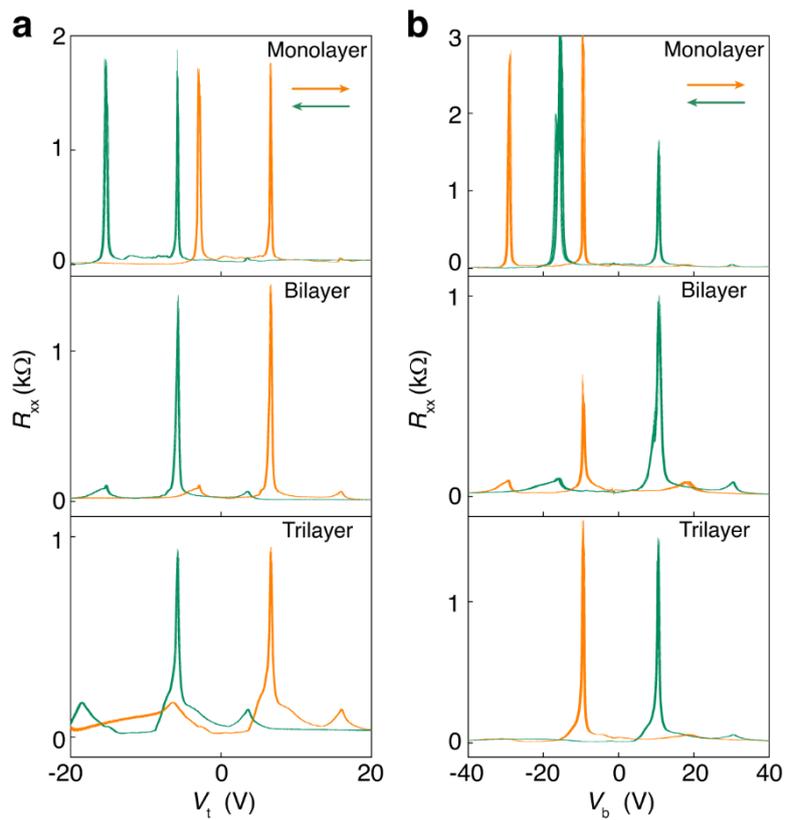

**Extended Data Fig. 4 | Robustness of hysteresis. a,b** Repetitive scan $V_t$ at a fixed $V_b = 0$ V (**a**) or scan $V_b$ at a fixed $V_t = 0$ V (**b**) for 20 circles in mono-/bi-/tri-layer graphene. All the curves overlap with each other, manifesting that the hysteresis behaviors are robust.

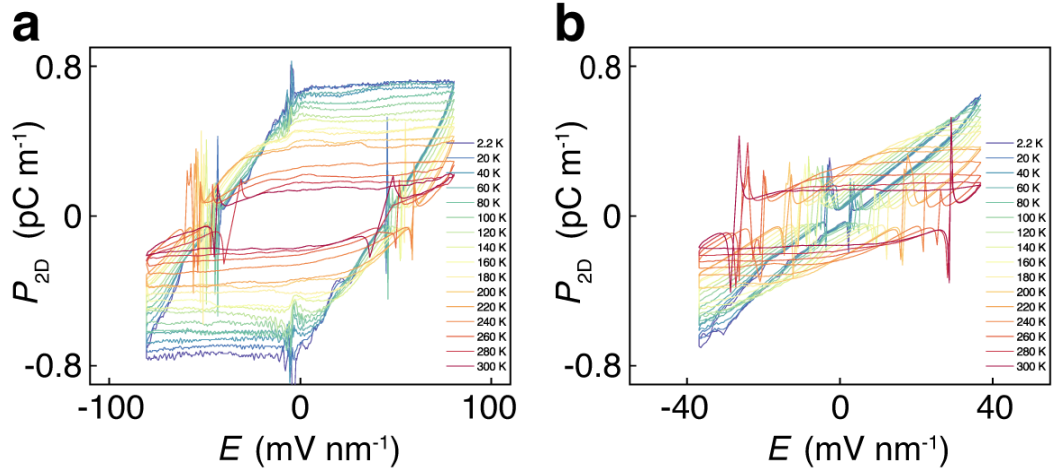

**Extended Data Fig. 5 | Temperature dependent $P_{2D} - E$ hysteresis loops. a,b,** $P_{2D} - E$ hysteresis loops measured at various temperature for two representatives $|E|_{max} = 80$ mV nm$^{-1}$ (**a**) and 36 mV nm$^{-1}$ (**b**). The temperature dependent $P_r$ shown in Fig. 3f were extracted from this figure. Additionally, we observed $P_s$ decreased with the increasing of temperature for both (**a**) and (**b**) and the windows of $P_{2D} - E$ hysteresis loops were broadened at elevated temperature.

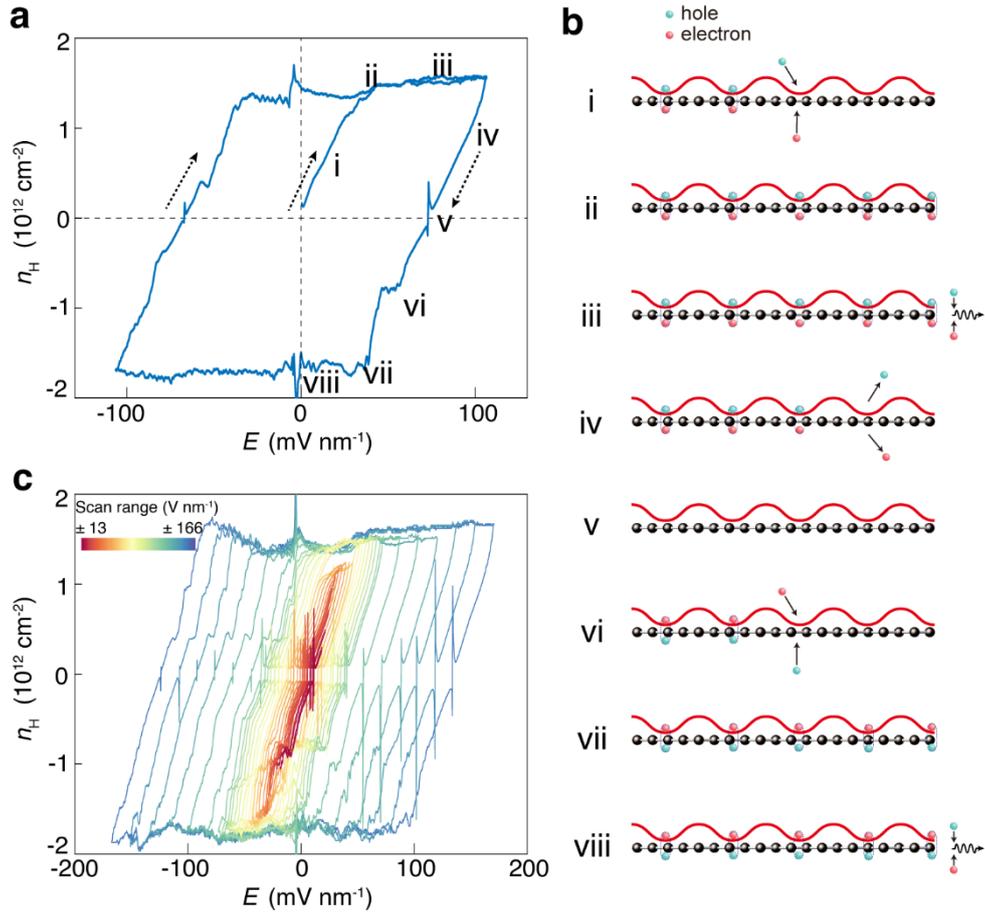

**Extended Data Fig. 6 | Untying the ferroelectric loops. a,** Hall carrier density $n_H$ as a function of $E$, which is re-plotted from $P_{2D} - E$ loop shown in Fig. 3b. One half cycle of the loop is divided into several processes. **b**, Schematic of charge polarization and saturation at each process marked in (**a**). The red wavy lines illustrate the moiré potential. The blue (red) ball denotes hole (electron). The arrows illustrate the injection (or extraction) process of charge carriers at top and bottom interfaces controlled by $V_t$ and $V_b$, respectively. The black wavy lines in Step iii and viii denote the recombination of electrons and holes. Step i, iv, v and vi are GSAS regions. Step ii, iii, vii, and viii are the normal dual-gating regions. **c**, Hall carrier density $n_H$ as a function of $E$ with a dense of scan range. The loops saturate quickly and broaden with further increasing of scan range. The data were acquired in monolayer graphene.

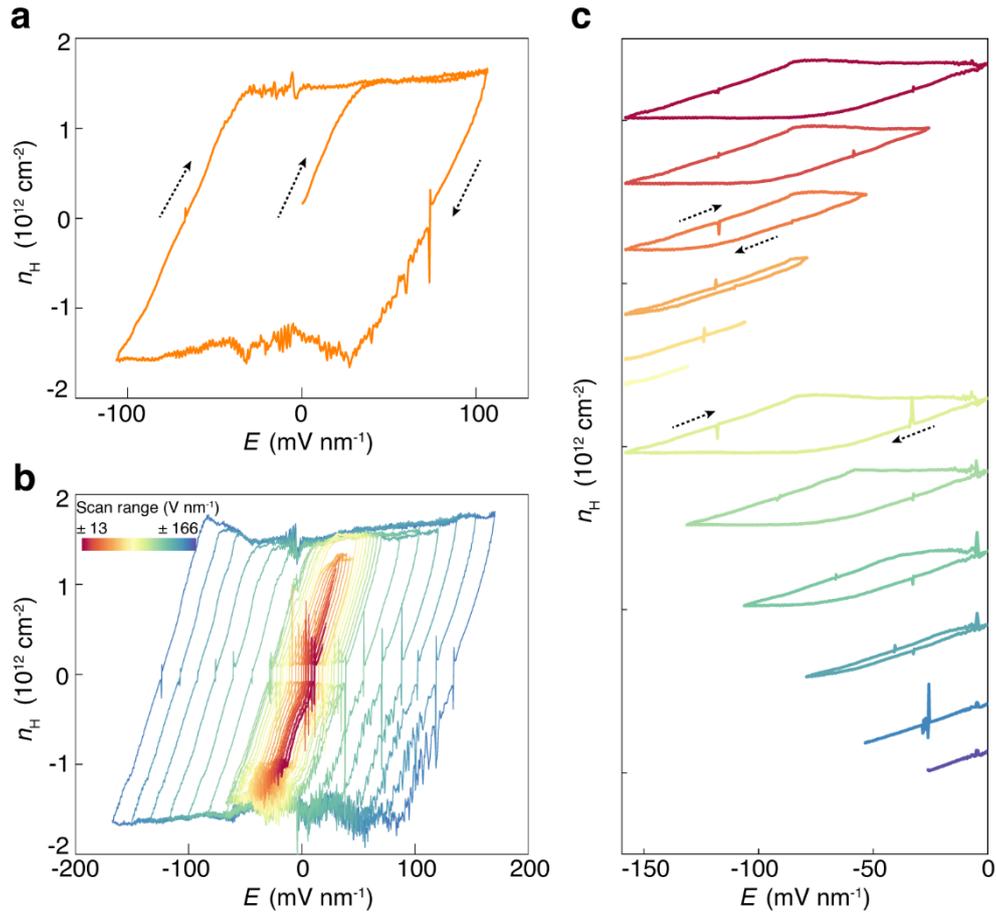

**Extended Data Fig. 7 | Ferroelectric loops in bilayer graphene. a**, Hall carrier density $n_H$ as a function of $E$ in bilayer graphene measured at $n_{total} = 0$. The electric field was scanned in the direction of $0 \rightarrow 106 \rightarrow -106 \rightarrow 106$ mV nm$^{-1}$ marked as black dash arrows. **b**, Ferroelectric loops for various scanning ranges. All the loops are similar to those of monolayer shown in Extended Data Fig. 6c. Particularly, the remanent Hall carrier density $n_H^s$ is the same as that in monolayer graphene, and approximately equal to $n_{Full}/2$. **c**, Ferroelectric loops in arbitrary scanning ranges.

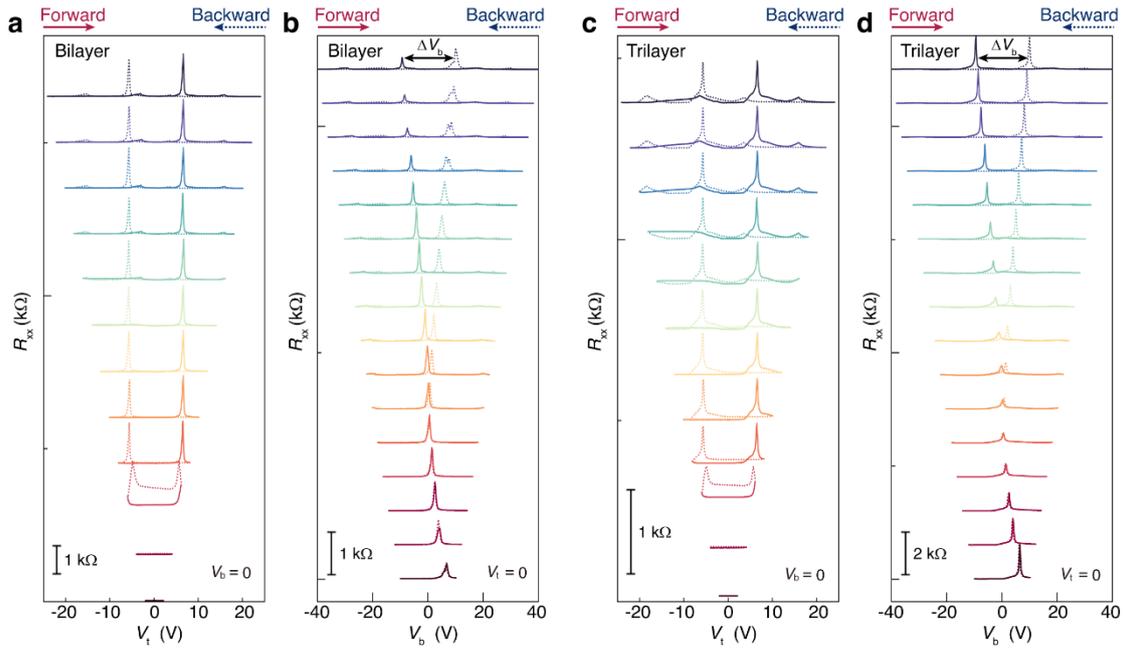

**Extended Data Fig. 8 | Ferroelectric hysteresis in bilayer and trilayer graphene. a**, $R_{xx}$ as a function of $V_t$ by sweeping $V_t$ in various ranges ($|V_t|_{max}$) while fixing $V_b = 0$. **b**, $R_{xx}$ as a function of $V_b$ by sweeping $V_b$ in various ranges ($|V_b|_{max}$) while fixing $V_t = 0$. The curves are vertically shift for clarity. (**a**) and (**b**) are acquired in bilayer graphene. **c,d**, Similar measurements to that in (**a**) and (**b**), but for trilayer graphene. Both bilayer (**a-b**) and tri-layer (**c-d**) graphene exhibit the same ferroelectric hysteresis as that in monolayer graphene presented in Fig. 1d-1e in the main text.

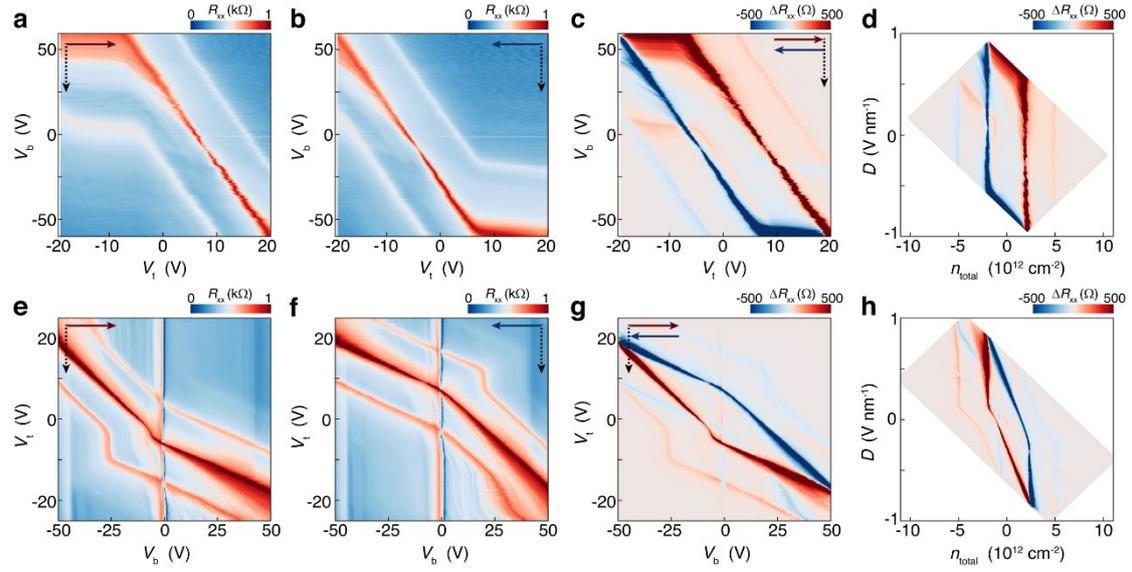

**Extended Data Fig. 9 | GSAS in bilayer graphene. a,b,** Dual-gate maps of $R_{xx}$ by scanning $V_t$ forward (**a**) and backward (**b**) at each fixed $V_b$. **c**, The difference between $R_{xx}$ in (**a**) and (**b**). **d**, The corresponding $n_{total} - D$ plot of (**c**). **e,f,** Dual-gate maps of $R_{xx}$ by scanning $V_b$ forward (**e**) and backward (**f**) at each fixed $V_t$. **g**, The difference between $R_{xx}$ in (**e**) and (**f**). **h**, The corresponding $n_{total} - D$ plot of (**g**). The data is collected simultaneously with that of Fig. 2 in the main text. The GSAS behavior in bilayer moiré structure is identical to that in monolayer counterpart.

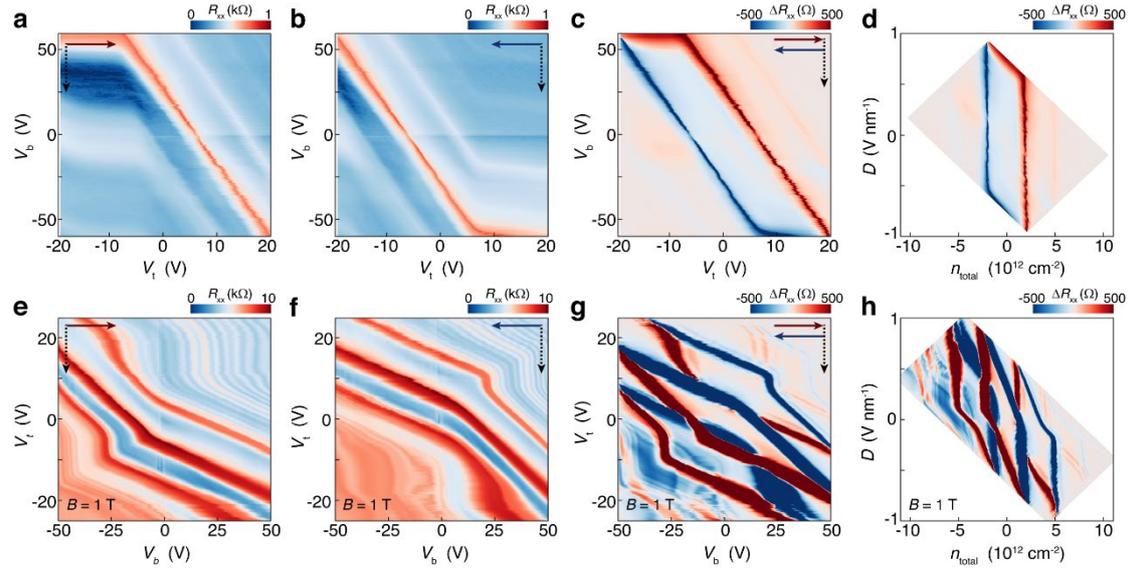

**Extended Data Fig. 10 | GSAS in trilayer graphene. a,b**, Dual-gate maps of $R_{xx}$ by scanning $V_t$ forward (**a**) and backward (**b**) at each fixed $V_b$ without magnetic field. **c**, The difference between $R_{xx}$ in (**a**) and (**b**). **d**, The corresponding $n_{total} - D$ plot of (**c**). **e,f**, Dual-gate maps of $R_{xx}$ by scanning $V_b$ forward (**e**) and backward (**f**) at each fixed $V_t$ with magnetic field $B = 1$ T. **g**, The difference between $R_{xx}$ in (**e**) and (**f**). **h**, The corresponding $n_{total} - D$ plot of (**g**). The electronic states of Landau levels also exhibit screening and hysteresis behaviors.